\documentstyle[12pt]{article}
\begin{document}

\hbadness=10000
\hbadness=10000
\begin{titlepage}
\nopagebreak
\def\thefootnote{\fnsymbol{footnote}}
\begin{flushright}
        {\normalsize
 DPSU-96-2\\
 LMU-TPW 96-3\\
January, 1996   }\\
\end{flushright}
\vspace{1cm}
\begin{center}
\renewcommand{\thefootnote}{\fnsymbol{footnote}}
{\large \bf  Soft Scalar Masses in String Models\\
with Anomalous $U(1)$ symmetry}

\vspace{1cm}

{\bf Yoshiharu Kawamura $^a$ 
\footnote[1]{e-mail: ykawamu@gipac.shinshu-u.ac.jp}}
and  \ {\bf Tatsuo Kobayashi $^b$
\footnote[2]{Alexander von Humboldt Fellow \\
\phantom{xxx}e-mail: kobayash@lswes8.ls-wess.physik.uni-muenchen.de}}

\vspace{1cm}
$^a$ Department of Physics, Shinshu University \\

   Matsumoto, 390 Japan \\
and\\
$^b$       Sektion Physik, Universit\"at M\"unchen \\

       Theresienstr. 37, D-80333 M\"unchen, Germany \\

\end{center}
\vspace{1cm}

\nopagebreak

\begin{abstract}
We obtain the low-energy effective theory 
from string models with anomalous $U(1)$ symmetry.
The feature of soft supersymmetry breaking 
scalar masses and some phenomenological implications
are discussed.
We show that it is, in general, difficult to keep 
the degeneracy and the positivity
of squared soft scalar masses at the Planck scale.
\end{abstract}

\vfill
\end{titlepage}
\pagestyle{plain}
\newpage
\section{Introduction}
\renewcommand{\thefootnote}{\fnsymbol{footnote}}

Superstring theories are powerful candidates for the unification
theory of all forces including gravity.
There are various approaches to explore 4-dimensional (4D) string models, 
for example, the compactification on Calabi-Yau manifolds\cite{CY}, 
the construction of orbifold models\cite{Orb,4DST} and so on.
The effective supergravity theories (SUGRAs) have been derived
based on the above approaches\cite{ST-SG}.
The structure of SUGRA\cite{SUGRA} is constrained
by considering field theoretical non-perturbative effects 
such as a gaugino condensation\cite{gaugino} 
and stringy symmetries such as duality\cite{duality}
besides of perturbative results.

Though the origin of supersymmetry (SUSY) breaking is unknown,
soft SUSY breaking terms have been derived under the assumption that
the SUSY is broken by $F$-term condensations of the dilaton field $S$
and/or moduli fields $T$\cite{ST-soft}.
Some phenomenologically interesting features are predicted from 
the structure of soft SUSY breaking terms which
are parameterized by a few number of parameters, for example,
only two parameters such as a goldstino angle $\theta$ and the gravitino 
mass $m_{3/2}$ in the case with the overall moduli and the vanishing 
vacuum energy\cite{BIM}.
Further cases with multimoduli fields are discussed 
in Refs. \cite{multiT}.

Most of 4D string models have anomalous $U(1)$ symmetries.
Some interesting features are pointed out in those models.
Fayet-Iliopoulos $D$-term\cite{FI} is induced at one-loop level
for anomalous $U(1)$ symmetry\cite{ST-FI}.\footnote{
Some conditions for absence of anomalous $U(1)$ are 
discussed in ref.\cite{anom}.}
As a result, some scalar fields necessarily develop vacuum expectation 
values(VEVs) and some gauge symmetries can break down\cite{ST-FI2}.

Such a symmetry breaking generates an intermediate scale $M_I$,
which is defined as the magnitude of VEVs of scalar fields,
 below the Planck scale $M_{Pl}$.
Using the ration $M_I/M_{Pl}$, higher dimensional couplings could explain 
hierarchical structures in particle physics like the fermion masses and 
their mixing angles.
Recently  much attention  has been paid to such a study 
on the fermion mass matrices \cite{texture,texture2}.
In Refs. \cite{texture}, $U(1)$ symmetries are used to generate realistic 
fermion mass matrices and some of them are anomalous, while stringy 
selection rules on nonrenormalizable couplings 
are used in Refs. \cite{texture2}.
Hence it is interesting to examine what features 4D string models 
with anomalous $U(1)$ symmetry can show at low energy or whether
we can construct a realistic model.

In this paper, we derive the low-energy theory 
from 4D string models with the anomalous $U(1)$ symmetry.
We discuss the feature of soft supersymmetry breaking scalar masses
and some phenomenological implications.
In particular,
we study the degeneracy and the positivity of squared scalar masses.
This subject has not been completely examined in the literatures
\cite{N,CD}.

\section{Structure of soft scalar mass}

Let us explain our starting point and assumptions first.
We assume that 4D string models are described as the effective
SUGRA at the Planck scale $M_{Pl}$.
The gauge group is $G=G'_{SM} \times U(1)_A$ where 
$G'_{SM}$ is a group which contains the gauge group of the 
standard model, 
$SU(3)_C \times SU(2)_L \times U(1)_Y$ as a subgroup
and $U(1)_A$ is anomalous.
This anomaly is canceled by the Green-Schwarz mechanism\cite{GS}.
The chiral multiplets are classified into two categories.
One is a set of $G'_{SM}$ singlet fields with large VEVs
denoted as $\Phi^i$.
It is assumed that the SUSY is broken by those $F$-term condensations.
Some of them have non-zero $U(1)_A$ charges and 
induce to the $U(1)_A$ breaking.
$S$ and $T$ belong to $\{\Phi^i\}$.
Here we treat only the overall moduli field $T$, but not 
several moduli fields.
The second one is a set of $G'_{SM}$ non-singlet 
fields $\Phi^{\kappa}$.
For simplicity, we treat all $\Phi^{\kappa}$'s as light fields
whose masses are small compared with $M_I$.
We denote the above two types of multiplet as $\Phi^I$ .

We study only a simple case with the following assumptions
to avoid a complication.
\begin{enumerate}
\item The $U(1)_A$ breaking scale is much higher than 
that of $G'_{SM}$.
We introduce one chiral matter multiplet $X$ with
a large VEV of order $M_I$ to break $U(1)_A$.

\item The VEV of $X$ is much smaller than those of $S$ and $T$, i.e.
\begin{eqnarray}
\langle X \rangle \ll \langle S \rangle, \langle T \rangle = O(M) ,
\end{eqnarray}
where $M$ is the gravitational scale defined as 
$M \equiv M_{Pl}/\sqrt{8\pi}$.
Hereafter we take $M = 1$.

\item Effects of threshold corrections and a $S$-$T$ mixing 
are small and neglected.
\end{enumerate}
It is straightforward to apply our method to more complicated 
situations.
We will comment on some of them later.

Our starting SUGRA is determined by the following three gradients,
that is, the K\"ahler potential $K$,
the superpotential $W$ and the gauge kinetic function $f_{\alpha}$.
Orbifold models lead to the following K\"ahler potential $K$: 
\cite{ST-SG,OrbSG}
\begin{eqnarray}
   K &=& -ln(S + S^* + \delta_{GS}^{A} V_A) - 3ln(T + T^*)
\nonumber\\ 
&~& + (T + T^*)^{n_X}|X|^2  + (T + T^*)^{n_\kappa}|\Phi^\kappa|^2 + \cdots,
\label{K}
\end{eqnarray}
where $\delta_{GS}^{A}$ is a coefficient of the Green-Schwarz
mechanism to cancel $U(1)_A$ anomaly and 
$V_A$ is a vector superfield of $U(1)_A$.
The dilaton field $S$ transforms nontrivially as 
$S \rightarrow S-i\delta_{GS}^{A}\theta(x)$ under $U(1)_A$ 
with the transformation parameter $\theta(x)$.
The coefficient $\delta_{GS}^{A}$ is given as 
\begin{eqnarray}
  \delta_{GS}^{A} &=& {1 \over 96\pi^2}Tr Q_A.
\label{delta_GS}
\end{eqnarray}
We estimate as $|\delta_{GS}^{A}/q_X| = O(10^{-1}) \sim O(10^{-3})$
by using explicit models.\footnote{
For example, see Refs.\cite{ST-FI2,LRmodel}.}
Here $q_X$ is a $U(1)_A$ charge of $X$.
And $n_I$'s are modular weights of matter multiplets $\Phi_I$. 
The same K\"ahler potential is derived from Calabi-Yau models 
with the large $T$ limit.
If the VEV of $X$ is comparable with one of $T$, we should replace 
the second and third terms in Eq.~(\ref{K}) as 
\begin{eqnarray}
-3ln(T + T^*-|X|^2),
 \end{eqnarray}
for the untwisted sector and  
\begin{eqnarray}
-ln [( T + T^*)^3-( T + T^*)^{n_X+3}|X|^2 ],
\end{eqnarray}
for the twisted sector.
The superpotential $W$ has $U(1)_A$ invariance.
We examine its consequence at low energy
without specifying the form of $W$ in this paper.
Note that the term dependent on only $X$ is forbidden 
by the $U(1)_A$ invariance.
The total K\"ahler potential $G$ is defined as $G \equiv K + ln|W|^2$.
The gauge kinetic function $f_{\alpha}$ is given as
$f_{\alpha} = S$.
For simplicity, here we assume the Kac-Moody levels satisfy 
$k_\alpha=1$ , 
because our results on 
soft terms are independent of a value of $k_\alpha$.
The scalar potential is given as
\begin{eqnarray}
   V &=& V^{(F)} + V^{(D)} ,
\label{V}\\
   V^{(F)} &\equiv& e^G(G^I (G^{-1})_I^J G_{J}-3),
\label{V(F)}\\
V^{(D)} &\equiv& {1 \over S+S^*} (D^A)^2 + {1 \over S+S^*} (D^a)^2
\nonumber\\
      &=& {1 \over S+S^*} ({\delta_{GS}^{A} \over S+S^*}
 + q_X K_X X + q_\kappa K_\kappa \Phi^\kappa)^2 
\nonumber\\
&~&+ {1 \over S+S^*} (K_{\kappa} (T^a \Phi)^{\kappa})^2 ,
\label{V(D)}
\end{eqnarray}
where the indexes $I$, $J$,... run all scalar species,
the index $a$ runs the generators of $G'_{SM}$ gauge group and
the $U(1)_A$ charge of matter multiplet $\Phi^I$ is denoted as $q_I$.
Note that the Fayet-Iliopoulos $D$-term\cite{FI} appears 
in $V^{(D)}$ for $U(1)_A$ if we replace $S$ by its VEV.
The $U(1)_A$ is broken by the condensations of $S$ and $X$.
The $U(1)_A$ breaking scale is of order $\langle N \rangle$
where $N$ is a Nambu-Goldstone multiplet.
The gravitino mass is obtained as $m_{3/2}^2=e^G$.
We assume that $V^{(F)} \leq O(m_{3/2}^2M^2)$.

Next we explain the procedure to obtain the low-energy
theory.
\begin{enumerate}
\item We write down the scalar potential $V$ by using the 
variations $\Delta \Phi^i = \Phi^i - \langle \Phi^i \rangle$.
We treat $\Phi^i$'s as dynamical fields.

\item We identify the Nambu-Goldstone multiplet $N$
related to $U(1)_A$ breaking whose mass is the same order of 
that of $U(1)_A$ gauge boson
by calculating the scalar masses.

\item Then we solve the stationary conditions of the potential 
for $\Phi^i$ while keeping the light fields arbitrary
and integrate out the heavy field $N$ by inserting the solutions 
into the scalar potential.
Simultaneously we take the flat limit, while fixing $m_{3/2}$ finite.
\end{enumerate}

We can obtain the scalar potential ${V}^{eff}$ of the effective theory
by the straightforward calculation\cite{Kawa}.
Here we write down the result in a model-independent manner as follows,
\begin{eqnarray}
{V}^{eff}&=& V_0 + {V}_{SUSY}^{eff} + {V}_{Soft}^{eff},
\label{Veff}\\
V_0 &=& \langle e^G(G^i (G^{-1})_i^j G_{j}-3) \rangle,
\label{V0}\\
{V}_{SUSY}^{eff} &=&
|\frac{\partial \hat{W}_{eff}}{\partial \Phi^{\kappa}}|^2
+ {1 \over 2}g_a^2 (K_{\kappa}(T^a \Phi)^{\kappa})^2,
\label{VSUSY}\\
{V}_{Soft}^{eff}&=& A \hat{W}_{eff} 
   + B^{\kappa}(\Phi) \frac{\partial \hat{W}_{eff}}{\partial \Phi^{\kappa}}
                + {\it H.c.}
\nonumber \\
&~&+ (m^2)_{\kappa}^{\lambda} \Phi^{\kappa} \Phi^*_{\lambda}
 + C_{\kappa\lambda} \Phi^{\kappa} \Phi^{\lambda} 
                + {\it H.c.},
\label{Vsoft}
\end{eqnarray}
where $g_a$'s are the gauge coupling constants of $G'_{SM}$.
Here we use the relation $\langle S \rangle = 1/g_{\alpha}^2$
and omit the terms whose magnitudes 
are less than $O(m_{3/2}^4)$.
Note that there is no $D$-term contribution on the cosmological 
constant $V_0$.
We do not write down the explicit forms for
the effective superpotential $\hat{W}_{eff}$, parameter $A$,
field-dependent functions $B^{\kappa}$ or $C_{\kappa\lambda}$
since it is irrelevant to the later discussions.\footnote{
Consult the reference \cite{Kawa} if necessary.}
We are interested in only {\it chirality-conserving} scalar mass
$(m^2)_{\kappa}^{\lambda}$ in this paper.
The formula is given as 
\begin{eqnarray}
(m^2)_{\kappa}^{\lambda} &=& 
(m_{3/2}^2 + V_0) \langle K_\kappa^\lambda \rangle 
+ \langle {F}^i \rangle \langle ({K}_{i\kappa}^{\mu} (K^{-1})^{\nu}_{\mu}
{K}^{j\lambda}_{\nu} - {K}_{i\kappa}^{j\lambda}) \rangle
\langle {F}^{*}_j \rangle
\nonumber\\
&~& + q_{\kappa} g_A^2 \langle D^A \rangle 
\langle K_\kappa^\lambda \rangle
\label{mass}\\
\langle D^A \rangle &=& 2M_{A}^{-2} 
\langle F^i \rangle \langle F^{*}_j \rangle 
 \langle (D^A)_i^j \rangle,
\label{<D>}
\end{eqnarray}
where $M_{A}$ is the mass of $U(1)_A$ gauge boson and 
$g_A$ is a gauge coupling constant of $U(1)_A$. 
The last term in Eq.~(\ref{mass}) is so-called $D$-term contribution
to the scalar masses\cite{D-term}.

We can apply the above result 
(\ref{Veff}) -- (\ref{<D>}) 
to the effective SUGRA defined by (\ref{K}) -- (\ref{V(D)}).
For the analysis of soft SUSY breaking parameters, 
it is convenient to introduce the following parameterization 
\begin{eqnarray}
&~&\langle e^{G/2} (K^S_S)^{-1/2} G^S \rangle 
 = \sqrt{3} C m_{3/2} e^{i\alpha_S} \sin\theta ,
\label{G^S}\\
&~&\langle e^{G/2} (K^T_T)^{-1/2} (G^T + (K^T_T)(K^T_X)^{-1} G^X)
 \rangle  = \sqrt{3} C m_{3/2} e^{i\alpha_T} \cos\theta ,
\label{G^T}
\end{eqnarray}
where $(K^i_j)$ is a reciprocal of $(K^i_j)^{-1}$.
The vacuum energy $V_0$ is written as
\begin{eqnarray}
V_0 &=& 3 (C^2 - 1) m_{3/2}^2 + V_0(X) ,
\label{V0-para}\\
V_0(X) &\equiv& 
\langle e^{G} ([(K^X_X)^{-1} - (K^T_T)|K^T_X|^{-2}]|G^X|^2 ) \rangle .
\label{V0(X)}
\end{eqnarray}
Since $C^2$ should be positive or zero,
we have a constraint $V_0(X) \leq 3m_{3/2}^2 +V_0$ from 
Eq.~(\ref{V0-para}).
In the case with $V_0=0$, it becomes as
\begin{eqnarray}
[(K^X_X)^{-1}- (K^T_T)|K^T_X|^{-2}]|G^X|^2 \leq 3.
\label{Con}
\end{eqnarray}
It gives a constraint on the VEVs of $X$ and $T$.
Further a larger value of $V_0(X)$ in the above region means 
$C \ll 1$.
Such a limit as $C \to 0$ corresponds to the \lq\lq moduli-dominated" 
breaking, that is, $F^S \ll 1$ and $F^T$ and $F^X$ contribute 
to the SUSY-breaking.
Note that this situation does not agree with the case of 
the moduli-dominated 
breaking without the anomalous $U(1)$ breaking $sin \theta \to 0$.

The stationary conditions lead to  $\langle D \rangle = O(m_{3/2}^2)$, 
i.e. 
$\langle \delta^A_{GS}/(S+S^*)+q_XK_XX \rangle =0$ up to such an order.
When we expand the scalar potential by using the variations for
$\Phi^i$, we can identify the variation of Nambu-Goldstone multiplet $N$ as
\begin{eqnarray}
\Delta N &=& {1 \over a}[{\Delta S+\Delta S^* \over 2 \langle S \rangle} 
+(1+{\langle K_{\xi \xi}\rangle \langle X \rangle ^2 \over 
\langle K_{\xi} \rangle})
{\Delta X+\Delta X^* \over \langle X \rangle} \nonumber\\
&~&+{\langle K_{\xi T} \rangle \over \langle K_{\xi} \rangle}
(\Delta T + \Delta T^*)],
\label{N}
\end{eqnarray}where 
$\xi = |X|^2$ and 
\begin{eqnarray}
a^2={1 \over 4 \langle S \rangle^2} 
+{1 \over \langle X \rangle^2}
(1+{\langle K_{\xi \xi} \rangle \langle X \rangle^2 \over 
\langle K_{\xi} \rangle})^2
+({\langle K_{\xi T} \rangle \over \langle K_{\xi} \rangle})^2.
\label{a^2}
\end{eqnarray}
This field $\Delta N$ has a heavy mass of order $O(|q_X \delta^A_{GS}|^{1/2}M)$,
and the other linear combinations of $S$, $T$ and $X$ have 
light masses.\footnote{
We can estimate the order of their masses as $O(m_{3/2})$\cite{Kawa}.}
The mass of $U(1)_A$ gauge boson $M_A$ is given as
\begin{eqnarray}
M_A^2 = 2g_A^2 ( {\delta_{GS}^A}^2 \langle K^S_S \rangle
 + q_X^2 \langle K^X_X \rangle |\langle X \rangle|^2 ) .
\label{M_A}
\end{eqnarray}
We can find that $|\delta_{GS}^A/q_X| \ll 1$ corresponds to 
$\langle X \rangle \ll \langle S \rangle, \langle T \rangle$.
In this limit, $\Delta N$ and $M_A$ are as follows,
\begin{eqnarray}
\Delta N = \Delta X + \Delta X^*, \quad 
M_A^2 = 2 g_A^2 q_X^2 \langle K^X_X \rangle |\langle X \rangle|^2 .
\label{M_Aagain}
\end{eqnarray}
Further $V_0(X)$ is negligible in this limit.
Thus the scalar mass is rewritten as
\begin{eqnarray}
m^2_{\kappa} &=&  m_{3/2}^2 + V_0 + (m_F^2)_\kappa + (m_D^2)_\kappa, 
\label{m_k}\\
V_0 &\equiv& 3 (C^2 - 1) m_{3/2}^2 ,
\label{V0-para2}\\
(m_F^2)_\kappa &\equiv& m_{3/2}^2 C^2 n_\kappa \cos^2\theta ,
\label{m_F}\\
(m_D^2)_\kappa &\equiv& m_{3/2}^2 {q_\kappa \over q_X} 
(1 - C^2 n_X \cos^2\theta + 6C^2 \sin^2\theta ) ,
\label{m_D}
\end{eqnarray}
where $(m^2)_{\kappa}^{\lambda} 
= m^2_{\kappa} \langle K_{\kappa}^{\lambda} \rangle$.
Note that our result is not reduced 
to that obtained from the theory with Fayet-Iliopoulos $D$-term, 
which is derived from the effective SUGRA by taking the flat limit first, 
even in the limit that $|\delta_{GS}^A/q_X| \ll 1$.
This disagreement originates from the fact that we regard 
$S$ and $T$ as dynamical fields, that is, we use the stationary conditions 
$\partial V/ \partial \Phi^i = 0$ to calculate $D$-term condensation.
This mass formula plays a crucial role in the following discussion.

\section{Phenomenological implications of soft scalar mass}

Now we shall discuss phenomenological implications of our results, 
especially the degeneracy and the positivity
of the squared soft scalar masses.
Hereafter we take $V_0 = 0$, i.e. $C^2=1$.
First we give a general argument by using the mass formula 
\begin{eqnarray}
m^2_{\kappa} = m_{3/2}^2[1+n_{\kappa} \cos^2 \theta
 +{q_{\kappa} \over q_X}(7-n_X\cos^2 \theta -6 \cos^2 \theta)] .
\label{mass2}
\end{eqnarray}
Note that the coefficient of $q_\kappa/q_X$ 
in Eq.~(\ref{mass2}) is sizable.
That could lead to a strong non-universality of soft 
scalar masses.\footnote{
Recently much work is devoted to phenomenological implications of 
the non-universality\cite{nonuni}.}

\makeatletter
\def\siml{\mathrel{\mathpalette\gl@align<}}
\def\simg{\mathrel{\mathpalette\gl@align>}}
\def\gl@align#1#2{\lower.6ex\vbox{\baselineskip\z@skip\lineskip\z@
 \ialign{$\m@th#1\hfill##\hfil$\crcr#2\crcr{\sim}\crcr}}}
\makeatother
\subsection{Degeneracy of soft scalar masses}

We obtain the difference of the soft masses as 
\begin{eqnarray}
{\Delta m^2 \over m_{3/2}^2}=\Delta n\cos^2 \theta 
+{\Delta q \over q_X}(7-n_X\cos^2 \theta -6 \cos^2 \theta) ,
\end{eqnarray}
by using Eq.~(\ref{mass2}).
The experiments for the process of flavor changing neutral current (FCNC)
 require that $\Delta m^2/m_{3/2}^2  \siml 10^{-2}$ 
for the first and the second families in the case with 
$m^2_{\tilde q} \sim O(1)$TeV\cite{FCNC}.
Hence we should derive $\Delta m^2/m_{3/2}^2  \approx 0$ 
within the level of $O(10^{-2})$.
Hereafter $a \approx 0$ denotes such a meaning.

If $\Delta q/q_X \approx 0$, we have 
$\Delta m^2 /m_{3/2}^2=\Delta n \cos^2 \theta$.
In this case the limit $\cos^2 \theta \rightarrow 0$ leads to 
$\Delta m \rightarrow 0$.
It corresponds to the dilaton-dominated breaking, 
where soft masses are universal \cite{Sdomi,BIM}.
Unless $\Delta q/q_X \approx 0$, 
we needs \lq \lq fine-tuning'' on the value of $\cos \theta$ 
as 
\begin{eqnarray}
\cos^2 \theta \approx {7 \over 6+n_X-q_X\Delta n/ \Delta q}.
\label{cos^2}
\end{eqnarray}
This \lq \lq fine-tuning'' is possible only in the case where\footnote{
In Ref.~\cite{Dudas}, the relation between modular weights and 
$U(1)$ charges is discussed as $q_X \Delta n / \Delta q = n_X$.
This relation does not satisfy Eq.~(\ref{Con2}).}
\begin{eqnarray}
1 + {\Delta n \over \Delta q}q_X \leq n_X.
\label{Con2}
\end{eqnarray}
Let us study the implication of Eq.~(\ref{cos^2}).
In the case with $\Delta n=0$, Eq.~(\ref{cos^2}) is reduced 
$\cos^2 \theta \approx 7 /( 6 +n_X)$.
Such a value of $\cos \theta$ is possible if $n_X \geq 1$.
Since such modular weights require at least two oscillators,
they are not obtained naturally\cite{mod-wei}.
If we take more natural value, e.g. $n_X=-1$,
Eq.~(\ref{Con2}) is reduced to $\Delta n q_X /\Delta q \leq -2$.

If Eq.~(\ref{cos^2}) is satisfied, 
the soft scalar mass is written as 
\begin{eqnarray}
m_\kappa^2=m_{3/2}^2[1+{7(n_\kappa-\Delta nq_\kappa/\Delta q) \over 
6+n_X-\Delta n q_X / \Delta q}].
\end{eqnarray}

\subsection{Positivity of squared soft scalar masses}

The condition for the positivity of $m^2_{\kappa}$ is written as 
\begin{eqnarray}
-n_\kappa+{q_\kappa \over q_X}(n_X+6) \leq 
(1+7{q_\kappa \over q_X}) \cos^{-2} \theta .
\label{Con-3}
\end{eqnarray}
If $1+7q_\kappa /q_X$ is positive, 
we can find a solution $\cos \theta$ of the above 
constraint for any $n_\kappa, n_X, q_\kappa$ and $q_X$.
On the other hand, if $1+7q_\kappa /q_X$ is negative, 
it leads to the following constraint;
\begin{eqnarray}
1+n_\kappa \geq {q_\kappa \over q_X}(n_X-1),
\end{eqnarray}
because $\cos^{-2} \theta \geq 1$.

Let us consider two extreme examples for the SUSY-breaking.
One is the case of dilaton-dominated breaking $(\cos \theta =0)$.
In this case we have 
\begin{eqnarray}
m_\kappa^2=m_{3/2}^2(1+7{q_\kappa \over q_X}).
\label{Sdom}
\end{eqnarray}
The positivity of $m^2_{\kappa}$ requires that $1+7q_\kappa/q_X \geq 0$.
The other is that of moduli-dominated breaking $(\cos^2 \theta =1)$.
In this case we have 
\begin{eqnarray}
m_\kappa^2=m_{3/2}^2[1+n_\kappa+{q_\kappa \over q_X}(1-n_X)].
\label{Tdom}
\end{eqnarray}
For example in the case with $n_\kappa=n_X=-1$, the positivity 
is realized only if $q_\kappa/q_X$ is positive.

In both cases of Eqs.~(\ref{Sdom}) and (\ref{Tdom}), 
the fields with $q_\kappa/q_X<0$ can easily have negative 
squared scalar mass of $O(m_{3/2}^2)$ at the Planck scale.
That implies that several fields develop VEV's and they trigger symmetry 
breakings.
We can show that there exist fields with $q_\kappa/q_X<0$ 
for each gauge group other than $U(1)_A$.
The reason is as follows.
Let us assume the gauge group is $U(1)_A \times \prod_\ell G_\ell$.
The Green-Schwarz anomaly cancellation mechanism requires that 
$C_{G_\ell}=\delta_{GS}^A k_{\ell}$ for any $\ell$, where $C_{G_\ell}$ is 
a coefficient of $U(1)_A \times G_\ell^2$ anomaly 
and $k_{\ell}$ is a Kac-Moody level of $G_\ell$.
Through the $U(1)_A$ breaking due to the Fayet-Iliopoulos $D$-term, 
the field $X$ develops its VEV. 
Here its charge should satisfy $q_X Tr Q_A < 0$ and $q_X C_{G_\ell} < 0$.
Each gauge group $G_\ell$ always has fields $\Phi^\kappa$
which correspond nontrivial representation on its group 
and whose $U(1)_A$ charges satisfy $q_X q_{\kappa} < 0$ 
because of $q_X C_{G_\ell} < 0$.
The $D$-term contribution on soft terms is very sizable.
That could naturally lead to $m_{\kappa}^2 < 0$ except a narrow region 
and cause  $G_\ell$ breaking.

Next we show that the scalar mass can be a source
to break $G'_{SM}$
by using the explicit model\cite{LRmodel}.
The model we study is the $Z_3$ orbifold model with 
a shift vector $V$ and Wilson lines $a_1$ and $a_3$ such as
$$V={1 \over 3}(1,1,1,1,2,0,0,0)(2,0,0,0,0,0,0,0),$$
$$a_1={1 \over 3}(0,0,0,0,0,0,0,2)(0,0,1,1,0,0,0,0),$$
$$a_3={1 \over 3}(1,1,1,2,1,1,1,0)(1,1,0,0,0,0,0,0).$$
This model has a gauge group as 
$$ G = SU(3)_c \times SU(2)_L \times SU(2)_R \times U(1)^7 
\times SO(8)' \times SU(2)'$$
and matter multiplets as 
\begin{eqnarray*}
{\rm U-sec.} &: \quad &3[(3,2,1)_0+(\bar 3,1,2)_0+(1,2,2)_0] \\
&~&+3[(8,2)'_6+(1,1)'_{-12}],
\end{eqnarray*}
\begin{eqnarray*}
{\rm T-sec.} &: \quad &9[(3,1,1)_4+(\bar 3,1,1)_4] 
+15[(1,2,1)_4+(1,1,2)_4]\\
(N_{OSC}=0)&~& +3(1,2,2)_4+
3[(1,2,1)(1,2)'_{-2}+(1,1,2)(1,2)'_{-2}]\\
&~&24(1,2)'_{-2}+60(1,1,1)_4+3(1,1,1)_{-8},
\end{eqnarray*}
$${\rm T-sec.} (N_{OSC}=-1/3):\quad 9(1,1,1)_4,$$
where the number of suffix denotes the anomalous U(1) charge 
and $N_{OSC}$ is the oscillator number.
This model has $Tr Q_A=864$.
Let us call the singlet fields in the untwisted sector $u$.
Such fields have no charge other than the anomalous $U(1)$ charge.

Suppose that a linear combination of $u$ develops VEV.\footnote{
This assumption should be justifed by solving
the stationary conditions, but here we take it
because non-pertabative effects in $W$ are not fully understood.}
Note that such a VEV does not break any gauge groups except the 
anomalous $U(1)_A$.
Our mass formula (\ref{mass2}) holds in this case.
In Table 1, we give a ratio $m_\kappa^2/m_{3/2}^2$ for all species
in two extreme cases, $\cos^2 \theta = 0$ and 
$\cos^2 \theta = 1$.
More than half of the fields acquire  negative squared masses and 
they could trigger a \lq \lq larger'' symmetry breaking 
including the dangerous color and/or charge symmetry breaking.
In addition we have a strong non-universality of soft masses, i.e.  
$m_\kappa^2=O(10m^2_{3/2})$ 
for some fields while $m_\kappa^2=O(m^2_{3/2})$ for others.
However, in this model, soft masses are degenerate for squarks 
and sleptons with same 
quantum numbers under $G_{SM}$
because they have same quantum
numbers under gauge group $G$ and same modular weights.

\section{Conclusions and discussions}

We study the effective field theory below the anomalous $U(1)$ 
symmetry breaking from the viewpoint of superstring theory.
The $D$-term contributions on soft scalar masses are obtained and 
parameterized in terms of the goldstino angle.
These contributions are very sizable.
We find that the $F$-term contribution from the difference among
modular weights and
the $U(1)_A$ $D$-term contribution to scalar masses
can destroy universality among scalar masses at $M$.
This non-degeneracy endangers the discussion of the suppression
of FCNC process.
On the other hand, the difference among $U(1)$ charges
is crucial for the generation of fermion mass hierarchy.
It seems to be difficult to make two discussions compatible.
As a byway, we can take a model that the fermion mass hierarchy is 
generated due to non-anomalous $U(1)$ symmetry
and the SUSY is broken by the dilaton $F$-term condensation.
For example, it is supposed that anomalies from the contributions of
MSSM matters are cancelled out by those of extra matters in such a model.
Moreover many fields acquire negative squared masses and 
they could trigger a \lq \lq larger'' symmetry breaking 
including the dangerous color and/or charge symmetry breaking.
This type of symmetry breaking might be favorable in the case where 
$G'_{SM}$ is a large group like a grand unified group.
These results are very useful for model building.

There can exist other non-universal contributions
to soft scalar masses in addition to $F$-term and $D$-term
contributions discussed in this paper.
We have not  discussed them since we need a model-dependent analysis
which is beyond the scope of this paper.
Here we only give brief comments.
On the breakdown of extra symmetries, there can exist
$D$-term contributions, which give soft scalar masses 
a further non-universality.
We can show that their contributions vanish for non-anomalous
symmetries in the limit of dilaton dominant SUSY breaking.
If there exist complex scalars whose masses are of order
of the symmetry breaking scales, their $F$-term contributions 
can appear after integrating out them.
Both of them should be considered at the symmetry breaking scales.

Further we have to take into account $T$-dependent threshold
corrections and the $S$-$T$ mixing at the one-loop level.
Our approach is available in this case, too.
Actually such corrections are rather small
for most of cases.

The moduli fields have a problem in string cosmology
because their masses are estimated as of $O(m_{3/2})$
and they weakly couple with 
the observable matter fields, i.e. through the gravitational 
couplings \cite{cosmo}.
They decay slowly to the observable matter fields.
That makes the standard nucleosynthesis dangerous.
In our model, some linear combinations of $S$, $T$ and $X$ remain light
whose $F$-terms are of $O(m_{3/2} M)$ and break the SUSY.
It is supposed that the couplings between such fields and observable
fields are strongly suppressed to guarantee the stability of weak scale.
Such a problem have to be considered for the light linear combinations, too.

\section*{Acknowledgments}
The authors are grateful to J.~Louis, H.~Nakano, N.~Polonsky, 
D.~Suematsu and M.~Yamaguchi for useful discussions.  
This work of Y.K. is supported by the Grant-in-Aid for
Scientific Research ($\sharp$07740212) from 
the Japanese Ministry of Education, Science and Culture.

\newpage

\begin{center}
{\Large Table 1 } 
\end{center}

\begin{tabular}{|c|c|c|c|c|c|}\hline
 &Rep.& \#& $q_\kappa$ & \multicolumn{2}{|c|}{$m_\kappa^2/m_{3/2}^2$}\\ 
& & & & $\cos^2 \theta=0$ & $\cos^2 \theta=1$\\ \hline 
U-sec. & $Q_L$ $(3,2,1)$ & 3 & 0 & 1 & 0 \\
       & $Q_R$ $(\bar 3,1,2)$ & 3 & 0 & 1 & 0 \\
      & $H$ $(1,2,2)$ & 3 & 0 & 1 & 0 \\
      & $(8,2)'$ & 3 & 6 & $-5/2$ & $-1$ \\ 
      & $u$ $(1,1)'$  & 3 & $-12$ & 8 & 2 \\ \hline
T-sec.  & $(3,1,1)$ & 9 & 4     &  $-4/3$  & $-5/3$ \\
$(N_{OSC}=0)$ &$(\bar 3,1,1)$ & 9 & 4     &  $-4/3$  & $-5/3$ \\
 & $L$ $(1,2,1)$ & 15 & 4     &  $-4/3$  & $-5/3$ \\
 & $R$ $(1,1,2)$ & 15 & 4     &  $-4/3$  & $-5/3$ \\
 & $(1,2,2)$ & 3 & 4     &  $-4/3$  & $-5/3$ \\
 & $(1,2,1)(1,2)'$ & 3 & $-2$     &  13/6  & $-2/3$ \\
 & $(1,1,2)(1,2)'$ & 3 & $-2$     &  13/6  & $-2/3$ \\
& $(1,2)'$ & 24 & $-2$     &  13/6  & $-2/3$ \\
 & $(1,1,1)$ & 60 & 4     &  $-4/3$  & $-5/3$ \\
 & $(1,1,1)$ & 3 & $-8$     &  17/3  & 1/3 \\ \hline
T-sec.  & $(1,1,1)$ & 9 & 4     &  $-4/3$  & $-8/3$ \\
$(N_{OSC}=1/3)$ & & &   &   &  \\ \hline
\end{tabular}

\section*{Table Captions}

\renewcommand{\labelenumi}{Table~\arabic{enumi}}
\begin{enumerate}
\item The particle contents and the ratios of $m_{\kappa}^2/m_{3/2}^2$.
In the third column, the degeneracy factors are shown.
We refer to the chiral multiplets as $Q_L$ for left-handed quarks,
$Q_R$ right-handed quarks, $H$ Higgs doublets,
$L$ for left-handed leptons and $R$ for right-handed leptons.
\end{enumerate}

\end{document}